# Single-crystal field-effect transistors based on copper phthalocyanine

R. Zeis, T. Siegrist and Ch. Kloc

.

Bell Laboratories, Lucent Technologies, 600 Mountain Ave., Murray Hill, NJ 07974, USA

**Copper phthalocyanine (Cu-Pc) single crystals were grown by physical vapor transport and field effect transistors (FETs) on the surface of these crystals were prepared. These FETs function as p-channel accumulation-mode devices. Charge carrier mobilities of up to 1 cm$^2$/Vs combined with a low field-effect threshold were obtained. These remarkable FET-characteristics, along with the highly stable chemical nature of Cu-Pc make it an attractive candidate for device applications.**

# 1. Introduction

Since the first paper on copper phthalocyanine (Cu-Pc) of de Diesbach and von der Weid in 1927 [1], extensive research has been carried out on this material. The outstanding chemical stability and strong blue dye properties of Cu-Pc resonate through numerous papers and reviews. Several hundred literature references and patents describe the significance of Cu-Pc in science and technology. Mostly, it has been used as paint, dye for textiles and plastics as well as ballpoint pen inks and printing inks. [2] Even food coloring with Cu-Pc was announced. [2] Recently, Cu-Pc has also been applied in chemical sensors [3] and optical data storage. [4]

The semiconducting behavior of metal phthalocyanines was already described as early as 1948 [5], but only recently have thin-film field-effect transistors based on Cu-Pc been considered as potential candidates for flexible electronics. However, the reported low thin film field-effect mobilities of alpha-Cu-Pc [6], much lower than amorphous silicon, have limited the use of this material for transistor applications. Moreover, the high chemical stability of Cu-Pc distinguishes this material from other high mobility organic semiconductors, like pentacene or rubrene, and stimulates research on improving the electrical properties of this compound. [7]. Additionally, the lack of reports on photochemical reactivity of Cu-Pc suggests that this material is suitable for light emitting diodes, organic lasers, or solar cell applications.

We assumed that the capabilities of Cu-Pc have not been well recognized since most research has been conducted on thin films, which crystallize in the alpha phase polymorph, and where disorder and grain boundaries mask the intrinsic semiconducting properties. To avoid grain boundaries and limit the concentration of impurities and defects, we decided to use Cu-Pc single crystals for evaluation of transport properties.

## 2. Experimental details

Cu-Pc single crystals were grown by horizontal physical vapor transport in a stream of argon gas. The details of the reactor and method have been presented previously. [8] The evaporating material Cu-Pc (Aldrich Chemical Co.) was heated to 500°C in the hot zone of a two-zone furnace. The second zone was held at room temperature. Cu-Pc single crystals spontaneously grew on a quartz tube wall between the zones as elongated dark violet needles that were several centimeters long and several hundred micrometers wide, and with thickness ranging from 10μm to 500μm.

The crystal structure of these needles was investigated using an Oxford-Diffraction Xcalibur-2 diffractometer, and rocking curves were determined using a triple-axis custom 4-circle diffractometer with copper radiation.

On the surface of fresh grown crystals a typical field effect transistor structure was produced using a technique described in Podzorov et al. [9] (Fig. 1). Source and drain contacts were painted with a water-based solution of colloidal graphite. The gate-insulating layer consisted of a 0.7-1.3 μm paryleneN thin film. Thickness of the films were determined with a profilometer. On top of the parylene layer, the gate electrodes were painted with the colloidal graphite between source and drain. The channel capacitance has been calculated from the thickness of the insulating layer and the tabulated dielectric constant of Parylene N. The transistor characteristic was measured using a test fixture connected to a HP 4145B semiconductor parameter analyzer. The measurements were performed in darkness and ambient air.

### 3. Results and Discussion

Cu-Pc, $CuN_8C_{32}H_{16}$ crystallizes in the beta form [10], with monoclinic unit cell parameters a =14.616(2)Å, b =4.8042(6)Å, c =17.292(3)Å, and β=105.39(2) Å, and space group $P2_1/n$, Z=2. This unit cell may be obtained from the original unit cell by the transformation (0,0,-1; 0,1,0;1,0,1)). The structure is shown in Fig. 2. The molecular packing produces two individual tilted stacks of Cu-Pc molecules running along the b-axis that are tilted against each other by 90 degrees. Rocking curves of the (-101) face using the (-404) reflection indicated a small mosaic spread of the order of 0.05 degrees. (Fig. 3)

Because the Cu-Pc single crystals form long needles, we measured the charge-transport properties in direction parallel to the long crystal axis (b-axis) on the (-101) face. This direction corresponds to the strongest overlap between π-orbitals of adjacent molecules. Figure 4 shows current-voltage (I-V) characteristics of a Cu-Pc single crystal device with a channel length L of approximately 380μm and width W of 100μm. The channel width was limited by the width of the crystals.

For small source-drain voltages ($V_{SD}$) the FET operates in the linear regime. If the source-drain voltage is increased, the gate field is no longer uniform and a depletion area is formed at the drain contact. Beyond a certain source drain voltage the current becomes saturated.

From the trans-conductance characteristic, we obtained the threshold voltage, $V_T$, around -6V, at $V_{SD}$=-40V (Fig. 5). Assuming that the density of electrically active traps is proportional to the charge needed to fill them, the density of the charged traps at the Cu-Pc/ Parylene interface is estimated 1 *$10^{11}$ cm$^{-2}$. The low threshold voltage and resulting low trap density indicates high quality of single crystals.

A negative field-effect onset (-2V) in p-type transistors indicates a "normally-off" FET. From this trans-conductance characteristic we also determined an on/off ratio of $10^4$. The sharpness of the field-effect onset is characterized by the subthreshold swing. Our Cu-Pc single crystal FETs exhibit a subthreshold swing (S) of S=2.6 V/decade, which is equivalent to a normalized subthreshold swing ($S_i$) of 7V·nF/decade·cm$^{-2}$. For pentacene single crystal FETs, we obtain a smaller value ($S_i$ = 3V·nF/decade·cm$^{-2}$) [11]. On the other hand, a normalized subthreshold swing for thin film field effect transistor based on Cu-Pc estimated from Zang et al. [12] is 4-times higher than our value. This also indicates the low defect concentration in the single crystal channel.

From the square root of the source drain current ($I_{SD}^{1/2}$) versus gate voltage ($V_G$) characteristics (Fig. 5), we extracted a field-effect mobility of 1cm$^2$/Vs. The field-effect mobility is estimated for a fixed source drain voltage ($V_{SD}$) of –40V in the saturation regime using Eg. (1)

$I_{SD}=(W/2L)\mu C_i(V_G-V_T)^2$ (1)

This value is an order of magnitude higher than reported by Zang et al. [12] for thin film devices having source-drain electrodes sandwiched between copper phthalocyanine (Cu-Pc) and cobalt phthalocyanine (Co-Pc).

It is worth noticing that a mobility of 1 cm$^2$/Vs is the highest measured in our study but that mobilities between 0.4 and 1 cm$^2$/Vs are routinely achieved on numerous crystals from numerous batches. This also indicates that substantially improved thin film FETs could be produced by using optimized thin film technology and by tuning the composition and structure of phthalocyanine compounds.

**Conclusions**

We have found that the room temperature Cu-Pc mobility is about 1 cm$^2$/Vs and the on/off ration is larger than $10^4$. These transistor parameters are comparable with widely used amorphous silicon and the highest hole mobilities reported for "conventional organic semiconductors" like tetracene (1.3 cm$^2$/Vs) [13] or pentacene (2.2 cm$^2$/Vs) [11] (2-2.5 cm$^2$/Vs) [14]. Taking into account the exceptional chemical stability of copper phthalocyanine, this semiconductor seems to be the material of a choice for field effect transistors in display applications and organic solar cells applications. In both these applications, the large amount of light emitted (display) or absorbed (solar cells) would require a very stable semiconductor, and arenes (which photodimerizes and oxidizes) [11] or rubrene (which photo-oxidizes) [15] do not satisfy this requirement.

**Acknowledgments**

We thank Prof. E. Bucher for his support and advice, and E. Willamson and C. Besnard for their help in preparing the manuscript. R. Zeis acknowledges the financial support from the Konrad Adenauer Foundation, the German Academic Exchange Service (DAAD) and the Landesgraduiertenfoerderung Baden-Wuertemberg. We acknowledge the support of the US Department of Energy under grant # 04SCPE389.# References

[1] H. de Diesbach, and E. von der Weid, *Helv. Chim. Acta*, 10, 886, 1927

[2] F. H. Moser and A.L. Thomas, *The Phthalocyanines* (CRC, Boca Raton, 1983)

[3] G.G. Fedoruk, D.I. Sagaidak, A. V. Misevich, A. E. Pochtenny, *Sensors and Actuators B* 48, 351, 1998

[4] T. Reinot, W. H. Kim, J. M. Hayes, G. J. Small, *J. Opt. Soc. Am.* B14, 602, 1997

[5] A. T. Vartanyan, *Zh. Fiz. Khim.* 22, 769, 1948

[6] Zh. Bao, A. J. Lovinger and A. Dodabalapur, *Appl. Phys. Lett*. 69 (20), 3066, 1996

[7] R.W. I. De Boer, M. E. Gershenson, A. F. Morpurgo and V. Podorov, *Phys. Status Solid A 201 (6)*, 1302, 2004

[8] R. A. Laudise, Ch. Kloc, P. G. Simpkins and T. Siegrist, *J. Cryst. Growth* 187, 449, 1998

[9] V. Podzorov, V. M. Pudalov, and M. E. Gershenson, *Appl. Phys. Lett*. 82, 1739, 2003

[10] R.P. Linstead and J.M Robertson, *J. Chem. Soc. London*, 1736, 1936 and *Strukturbericht 4*, 320, 1936

[11] L. Roberson, J. Kowalik, L. Tolbert, Ch. Kloc, R. Zeis, X. Chi and C. Wilkins to be published

[12] J. Zhang, J. Wang, H. Wang, and D. Yan, *Appl. Phys. Lett*. 84, 142, 2004

[13] C. Goldmann, S. Haas, C. Krellner, K. P. Pernstich, D. J. Grundlach and B. Batlogg cond-mat/ 0403210, 2004

[14] V. Y. Butko , D. V. Lang , X. Chi , J. C. Lashley and A. P. Ramirez to be published

[15] V. Podzorov, V. M. Pudalov, and M. E. Gershenson cond-mat/ 0406738, 2004
7

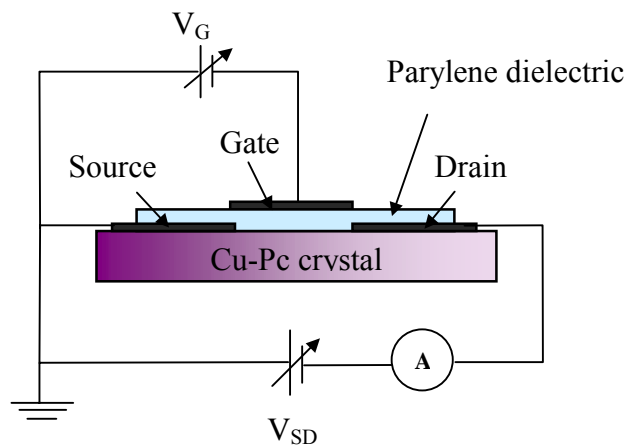

**Fig. 1.** Schematic representation of a Cu-Pc-based field effect transistor and the measuring circuit.

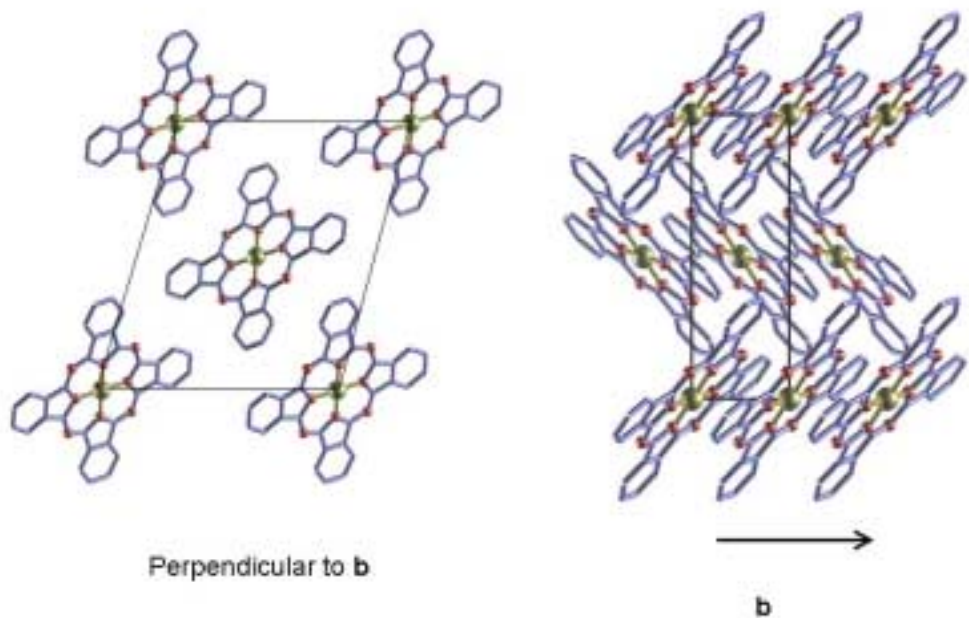

**Fig. 2.** The crystal structure of the ß phase of copper phtalocyanine (Cu-Pc). A strong π-orbital overlap exist along the b-axis.

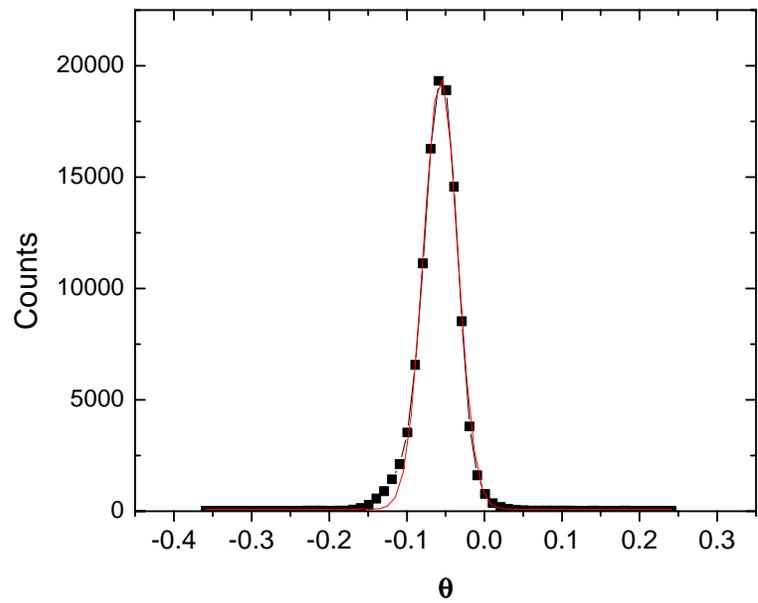

**Fig. 3.** The rocking curve of the Bragg reflection (-4 0 4).

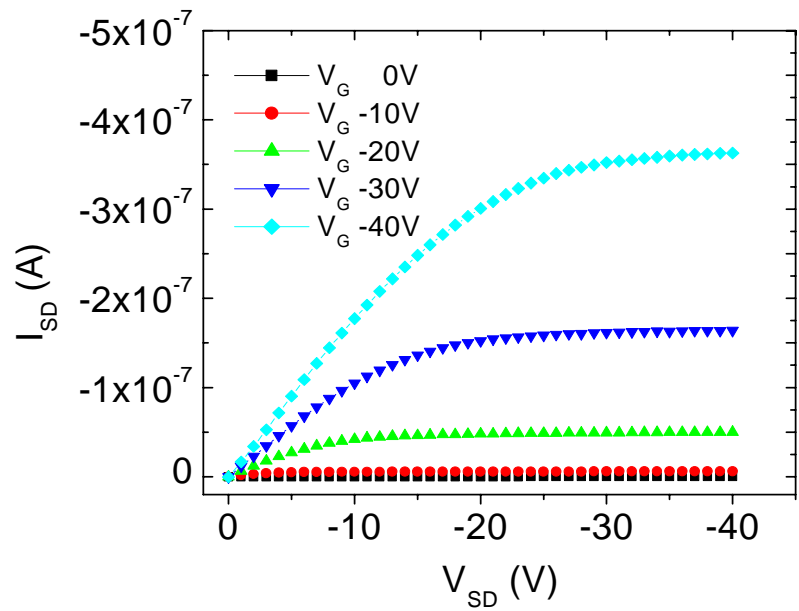

**Fig. 4.** The output characteristic of a Cu-Pc single crystal FET.

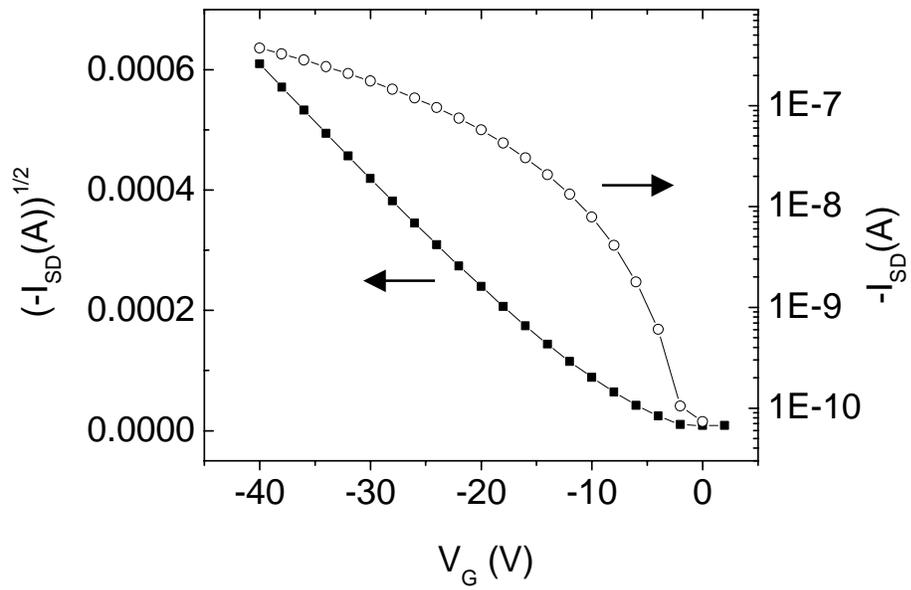

**Fig. 5.** The trans-conductance characteristic of a Cu-Pc FET measured at a fixed $V_{SD}=-40V$ (right axis) and the square root of the drain current in the saturation regime as a function of the gate voltage. (Left axis)